\documentclass[twocolumn,showpacs,preprintnumbers,amsmath,amssymb]{revtex4}
\usepackage{array}
\usepackage{booktabs}
\usepackage{tabu}
\usepackage{dcolumn}
\usepackage{amsmath}
\usepackage{amsfonts}
\usepackage{amssymb}
\usepackage{graphicx}
\usepackage{subfigure}
\usepackage{graphicx}% Include figure files
\usepackage{dcolumn}% Align table columns on decimal point
\usepackage{bm}% bold math

\begin{document}

%\preprint{APS/123-QED}

\title{Brane inflation driven by noncanonical scalar field}% Force line breaks with \\

\author{$^{a}$T. Golanbari}
 \email{t.golanbari@uok.ac.ir/@gmail.com}
 \author{$^b$A. Mohammadi}
  \email{abolhassanm@gmail.com}
  \author{$^a$Kh. Saaidi}
  \email{ksaaidi@uok.ac.ir;khaledsaeidi@gmail.com}
\affiliation{
$^a$Department of Physics, Faculty of Science, University of Kurdistan, Pasdaran St., P.O. Box 66177-15175, Sanandaj, Iran.\\
$^b$Young Researchers and elites Club, Sanandaj Branch, Islamic Azad University, Sanandaj, Iran.}
%This line break forced with \textbackslash\textbackslash

%\author{Charlie Author}
% \homepage{http://www.Second.institution.edu/~Charlie.Author}
%\affiliation{
%Second institution and/or address\\
%This line break forced% with \\
%}

\date{\today}% It is always \today, today,
             %  but any date may be explicitly specified

\def\br{\biggr}
\def\bl{\biggl}
\def\Br{\Biggr}
\def\Bl{\Biggl}
\def\be{\begin{equation}}

 \def\ee{\end{equation}}
\def\bea{\begin{eqnarray}}
\def\eea{\end{eqnarray}}
\def\f{\frac}
\def\n{\nonumber}
\def\l{\label}

\begin{abstract}
In this work, we are going to study the inflationary era of the Universe evolution by using second type of Randall-Sundrum model (RS-II) braneworld gravity. It is supposed that the Universe is dominated by scalar field with noncanonical kinetic terms. The kinetic term is supposed as a power-law function, and the work is performed for two typical cases. Using recent observations, the free parameters of the model are determined. It is shown that theoretical results are in acceptable agreement with observational data. Finally the time period of inflation is derived approximately, and it is found out that the inflation could occur after the five-dimensional Planck time.
\end{abstract}

%\pacs{Valid PACS appear here}% PACS, the Physics and Astronomy
                             % Classification Scheme.
\pacs{04.50.+h; 98.80.Cq; 98.80.Es; 98.80.Bp}
%\keywords{noncanonical scalar field; chaotic inflation; scalar and tensor perturbation.}%Use showkeys class option if keyword
                              %display desired
\maketitle

\newpage

%==========================================================================
%==========================================================================
%==============Introduction I =================================================
%==========================================================================
%==========================================================================
\section{Introduction}
Today, the Universe undergoes an accelerated expansion that is related to an ambiguous kind of matter known as dark energy. There is various evidence to confirm this expansion such as type Ia supernovae data \cite{1},
the Wilkinson microwave anisotropic probe (WMAP) \cite{2}, x rays \cite{3}, and large scale structure \cite{4} and it known that this unseen fluid fills about $73\%$ of whole Universe. Up to now, different proposals have been presented to describe dark energy as a cosmological constant \cite{5}, quintessence \cite{6}, k essence \cite{7}, phantom \cite{8}, modifying gravity \cite{9}, and so on , and it is well known that the scalar field plays an important leading role for most of these models. The type of expansion and formation of the Universe is strongly dependent on its early times evolution. It is strongly believed that early times of the Universe evolution could be described by the inflationary scenario. Based on this scenario, the scalar field dominates the Universe and causes a quasi-de Sitter expansion. Therefore, the Universe experiences an extreme expansion in a short period of time. Most of the scalar field models contain a canonical kinetic energy density. However, recently, the models of a scalar field with noncanonical kinetic energy density received huge attention. The general form of the Lagrangian for these kind of models is given by $L=f(\phi)F(X)-V(\phi)$, where $X=(\nabla_\mu\phi\nabla^\mu\phi) / 2$ \cite{10}. By taking $V(\phi)=0$, the Lagrangian comes into the well-known model named k essence. The idea of k essence was motivated from the Born-Infeld action of string theory \cite{11}, and it was first introduced as a possible model for inflation in Ref. \cite{12}. k essence comes to some interesting results about dark energy \cite{13}. For our present paper, we take $f(\phi)=1$, which expresses that the model is not exactly k essence. It describes a purely kinetic k essence including a potential term. This kind of model is known as a \textit{noncanonical scalar field} \cite{14}, which ie another class of  the general case. This class could be as important as  the k essence model and result in some interesting consequences. The cosmological solution of this model is considered in Ref. \cite{14}. They showed that, with a simple form of the $F(X)$ function, it is possible to build a unified model of dark energy and dark matter. In Ref. \cite{15}, the same case was studied, in which the authors found out that it is difficult to produce a unified model of dark energy and dark matter for purely kinetic k essence. Moreover, it was shown that the model is able to generate inflation in early times. These abilities of a noncanonical scalar field motivate us to study the model in more detail and for more scenarios. Similar to k essence, it is seems that this model has enough merit to be considered. Among the modified gravity theories, braneworld gravity has a special place. The model presents different picture of the Universe that has received huge attention. According to braneworld gravity, our Universe is a four-dimensional hypersurface (the brane) that is embedded in five-dimensional space-time (the bulk). All standard particles and their interaction are confined to the brane, and gravity could propagate along the fifth dimension. RS-II \cite{16} is one of the most important models of braneworld gravity. The model brings some modified terms for the Friedmann equation such as a quadratic of energy density, which could be very important in high-energy regimes (like the inflation epoch). However, the Friedmann equation comes back to standard four dimensions in the low-energy limit. \\
In the present paper, we are going to use a RS-II braneworld framework to study inflation governed by a noncanonical scalar field. The scenario of inflation in braneworld RS-II was considered in Ref. \cite{17}, where a usual canonical scalar field was used.  The authors showed that the inflationary scenario in the braneworld model could be more efficient than four-dimensional cosmology, and it could occur for wider ranges of potential. During this work, we use a common form of chaotic potential. To obtain the value of some important parameters of the scenario, such as spectra indices, we need to determine the free parameter of the model. It is performed by using recent observational data for the amplitude of scalar perturbation. Then, the theoretical results are compared with their corresponding observational values, and it is found out that they are in good agreement with each other. The paper is organized as follows. In Sec. II, the general form of the basic dynamical equation is derived. We assume a scalar field with a noncanonical kinetic term, and insert it into the evolution equation in Sec. III. Then, we impose slow-roll conditions on the dynamical equation. To investigate the work in more detail, we take the noncanonical kinetic term as a power-law function of $X$ in Sec.IV. The main equation is rewritten and the general form of slow-roll parameters is derived. The gauge-invariant curvature perturbation is obtained in next step, and final results are studied for two specific cases. In Sec.V, we summarize and discuss the results of the work.
%============================================================================
%============================================================================
%================= Section II ===================================================
%============================================================================
%============================================================================
\section{Framework and Basic Equations}\label{Sec1}
To begin, the action is assumed as following the usual form,
\begin{equation}\label{II.01}
S_5 = \int d^5x \sqrt{-g} \Big( {M_5^3 \over 2}\mathcal{R} - \Lambda \Big) + \int d^4x \sqrt{-q} \Big( L_{\rm b} - \lambda \Big),
\end{equation}
where $\mathcal{R}$ is a five-dimensional Ricci scalar, which is related to the five-dimensional metric $g_{AB}$. The bulk is filled with five-dimensional cosmological constant $\Lambda_5$. The five-dimensional Planck mass is denoted by $M_5$. The induced metric on the brane, $q_{\mu\nu}$, is related to five-dimensional metric $g_{AB}$ by $g_{AB}=q_{AB} + n_A n_B$, where $n^A$ is a unit normal vector. The Lagrangian of confined matter on the brane is described by $L_{\rm b}$, and $\lambda$ is the brane tension.

Taking a variation of action and using Z$_2$ symmetry leads one to the field equation for the brane \cite{18},
\begin{equation}\label{II.02}
G_{\mu\nu} = -\Lambda_4 g_{\mu\nu} + \left({8\pi\over M_{4}^2}\right) \tau_{\mu\nu}  + \left({8\pi\over
M_5^3}\right)^2 \Pi_{\mu\nu} - E_{\mu\nu}\ ,
\end{equation}
where
\begin{eqnarray}
\Lambda_4 & = & {4\pi \over M_5^3} \left( \Lambda + {4\pi\over3M_5^3} \lambda^2 \right),  \nonumber \\
E_{\mu\nu} & = & C_{MRNS}~n^M n^N q_{~\mu}^{R}~ q_{~\nu}^{S} , \nonumber \\
\tau_{\mu\nu} & = & -2 {\delta L_{\rm b} \over \delta g^{\mu\nu}} + g_{\mu\nu}L_{\rm b} , \nonumber \\
\Pi_{\mu\nu} & = & -\frac{1}{4} \tau_{\mu\alpha}\tau_\nu^{~\alpha} +\frac{1}{12}\tau\tau_{\mu\nu} +\frac{q_{\mu\nu}}{8}\tau_{\alpha\beta}\tau^{\alpha\beta} - \frac{q_{\mu\nu}}{24} \tau^2  . \nonumber
\end{eqnarray}
$\Lambda_4$ is an effective four-dimensional cosmological constant of the brane, and, according to Randall-Sundrum (RS) fine-tuning, could be set to zero. $E_{\mu\nu}$ is the projection of the Weyl tensor on the brane. The brane energy-momentum tensor is indicated by $\tau_{\mu\nu}$, and finally  $\Pi_{\mu\nu}$ is a contribution of the quadratic of the energy-momentum tensor in the field equation.

By the homogeneity and isotropy assumptions for the Universe, the five-dimensional Friedmann–Lemaitre–Robertson–Walker (FLRW) metric is introduced to describe the geometry of the Universe,
\begin{eqnarray} \label{II.03}
ds^2_5 = -dt^2 + a^2 \delta_{ij} dx^i dx^j + dy^2  ,
\end{eqnarray}
where $\delta_{ij}$ is a maximally symmetric three-dimensional metric, and $y$ is the fifth coordinate. According to observational data, the Universe is spatially flat in good approximation; then, we assume $k=0$
\footnote{Note that, in this paper, we are going to study the inflationary era of the Universe evolution, in which the Universe experiences huge expansion. Spatially curvature $k$ contributes to the Friedmann equation by the term $k/a^2$, which decreases as $a^{-2}$. Therefore, during inflation, this term is negligible, and ignoring it is a usual assumption in inflationary studies.}.
Substituting this metric in the field equations leads one to the following evolution equation:
\begin{equation}
H^2={\Lambda_4\over3} + \left({8\pi \over 3M_4^2}\right) \rho + \left({4\pi\over 3M_5^3}\right)^2\rho^2+{\mathcal{C} \over a^4} . \nonumber
\end{equation}
$\mathcal{C}$ is arising from $E_{\mu\nu}$ and describes the influence of bulk the graviton on brane evolution. This term is known as \textit{dark radiation}, which rapidly diluted during inflation \cite{17}. Therefore, one can ignore it and rewrite the Friedmann equation as
\begin{equation}\label{II.04}
H^2 = {8\pi \over 3M_4^2} \rho \left( 1+ {\rho \over 2\lambda} \right) ,
\end{equation}
where the four-dimensional Planck mass is defined as $M_4^2 = {3 \over 4\pi} {M_5^6 \over \lambda}$.
At  the low-energy density regime, where $\rho \ll \lambda$, the standard Friedmann equation is recovered. However, at the high-energy regime, $\rho \gg \lambda$, the quadratic term of the energy density plays an important role, so that it overcomes the linear term completely. On the other hand, the standard cosmology could describe nucleosynthesis successfully. Therefore, the modified Friedmann equation (\ref{II.04}) must come back to standard form at the nucleosynthesis time. In other words, the term $\rho^2$ should be subdominated at nucleosynthesis. This point results in a constraint for brane tension as $\lambda \geq 1 {\rm MeV^4}$; then, $M_5 \geq 10 {\rm TeV}$ \cite{19}. In the RS braneworld model with an infinite fifth dimension, the correction to the Newtonian law of gravity is of order $M_5^6 / \lambda^2 r^2$. This correction should be small on scales $r \geq 1 {\rm mm}$, which becomes a stronger constraint as $M_5 > 10^5 {\rm TeV}$ \cite{16}. In the present work, we assume that inflation occurs at the high-energy limit $\rho \gg \lambda$ \cite{20}, in which the quadratic term could overcome the linear term \cite{17,21,22}. Then, the Friedmann
equation could be written as
\begin{equation}\label{II.05}
H^2 =\left({4\pi \over 3M_5^3}\right)^2 \rho^2 .
\end{equation}
The time derivative of  the Hubble parameter is expressed by
\begin{equation}\label{II.06}
\dot{H} = - 3\left({4\pi \over 3M_5^3}\right)^2 \rho (\rho + p) .
\end{equation}
To derive the above relation, we use the conservation equation, which is given the same as standard cosmology as $\dot\rho + 3H(\rho + p) = 0$.
The acceleration of the Universe is obtained from Eqs.\;(\ref{II.05}) and (\ref{II.06}) by the following:
\begin{equation}\label{II.07}
{\ddot{a} \over a} = H^2 + \dot{H} = - \left({4\pi \over 3M_5^3}\right)^2 \rho (2\rho + 3p) .
\end{equation}
Inflation is an era in which the Universe expands with positive acceleration; therefore, in Eq.\;(\ref{II.07}),
the acceleration condition $ p < -{2 \over 3}\rho $ should be satisfied. In comparison to standard cosmology, there is a
difference in the acceleration condition. The difference is resulted from a different Friedmann equation. \\
%============================================================================
%============================================================================
%================= Section III ==================================================
%============================================================================
%============================================================================
\section{NonCanonical Scalar Field}
In inflation studies, it is usually supposed that the Universe is dominated by a canonical scalar field that describes inflation. In this model, it is assumed that the matter Lagrangian is described by a noncanonical
scalar field that leads to a huge
expansion in early times of the Universe evolution (inflation). The noncanonical scalar field Lagrangian is denoted by the relation
\begin{equation}
L_b = F(X) - V(\phi) ,  \nonumber
\end{equation}
where $\phi$ is a scalar field and $V(\phi)$ is a self-interacting potential of the scalar field. $X$ is defined as $ X \equiv -(q^{\mu\nu}\nabla_{\mu}\phi \nabla_{\nu}\phi)/2 $, and $F(X)$ is an arbitrary function of $X$. Then, the energy density and pressure could be found out, respectively, as
\begin{eqnarray}\label{III.8}
\rho = 2XF_X - F + V, \qquad p = F - V ,
\end{eqnarray}
in which $F_X$ indicates the derivative of $F$ with respect to $X$. Substituting the above energy-momentum components  in
Eqs.\;(\ref{II.05}) and (\ref{II.06}), we can easily derive the evolution equations. The condition for having an acceleration phase is described by $V > 4XF_X + F$. On the other hand, the conservation equation could be expressed in a different way as
\begin{equation}\label{III.9}
(2XF_{XX} + F_X) \ddot{\phi} + 3F_X H \dot\phi + V' =0 ,
\end{equation}
which is known as the scalar field equation of motion (or wave equation) and governs on the scalar field evolution.\\
Based on Ref. \cite{23}, to have a quasiexponential expansion, the rate of the Hubble parameter during a Hubble time should be much smaller than unity, namely, $({|\dot{H}| / H^2}) \ll 1$. The same argument is supposed for $\dot\phi$, which leads to the condition $({|\ddot{\phi}| / H |\dot\phi|}) \ll 1$.
These two conditions are known as slow-roll approximations. Slow-rolling parameters are given by \cite{23}
\begin{equation}\label{III.10}
\epsilon_H = -{\dot{H} \over H^2}, \qquad \eta_H = - {|\ddot{\phi}| \over H |\dot\phi|} .
\end{equation}

To have inflation, the slow-roll parameters should be smaller than unity, namely $\epsilon_H, |\eta_H| \ll 1$. The smallness of $\epsilon_H$ lets us to ignore the kinetic term of the scalar field in the Friedmann equation against the potential term, and the smallness of $\eta_H$ lets us ignore $\ddot\phi$ against $\dot\phi$ in Eq.\;(\ref{III.9}). Imposing these conditions on dynamical equations leads one to
\begin{eqnarray}
H & = & {4\pi \over 3M_5^3} V(\phi) , \label{III.11}\\
\dot{H} & = & -3\left({4\pi \over 3M_5^3}\right)^2 V(\phi) \big( 2XF_X \big) , \label{III.12}\\
{\ddot{a} \over a} & = & \left({4\pi \over 3M_5^3}\right)^2 V^2(\phi) . \label{III.13}
\end{eqnarray}
Positivity of acceleration is clear from Eq.\;(\ref{III.13}). Also, for the scalar field equation of motion we have
\begin{equation}\label{III.14}
3F_X H\dot\phi + V'(\phi) = 0 .
\end{equation}
To go further, and consider the equations in more detail, we need to specify the kinetic term $F(X)$. In the next section, we shall pick out a power-law function for $F(X)$ and derive the important parameters of inflation.
%=============================================================================
%=============================================================================
%================= Section IV ===================================================
%=============================================================================
%=============================================================================
\section{Typical Examples}
In this section, we are going to consider the inflationary scenario by using a noncanonical scalar field and assuming a power-law function for $F(X)=F_0 X^n$, in which $F_0$ is a constant and its dimension is $ M^{4(1-n)}$. First, the general form of the equation is derived for this case. \\
Based on Eq.\;(\ref{III.8}), the energy density and pressure are expressed by
\begin{equation}\label{IV.15}
\rho = (2n-1)F_0 X^n + V(\phi), \qquad p = F_0 X^n - V(\phi).
\end{equation}
Using a power-law kinetic term in Eqs.\;(\ref{III.11}), (\ref{III.12}), and (\ref{III.14}), the general form of slow-rolling parameters is read as
\begin{eqnarray}
\epsilon_V & = & \left( {M_5^3 \over 4\pi n F_0\sqrt{2}} \right)^{ {2n \over 2n-1} } { 6nF_0 \over V(\phi)}
\left( -V'(\phi) \over V(\phi) \right)^{2n \over 2n-1},  \label{IV.16}\\
\eta_V & = & \left( {M_5^3 \over 4\pi n F_0\sqrt{2}} \right)^{ {2n \over 2n-1} } {6nF_0 \over 2n-1} {V''(\phi) \over V'^2 (\phi)} \left( -V'(\phi) \over V(\phi) \right)^{2n \over 2n-1}. \qquad \label{IV.17}
\end{eqnarray}
Up to now, the general form of dynamical equations and slow-rolling parameters is derived. To determine the free parameters of the model, we use the observational data. One of the common parameters is the amplitude of scalar perturbation, which is related to gauge-invariant curvature perturbation. In the following subsection, we briefly introduce this gauge-invariant parameter.
%============================================================================
%================= Subsection IV.A ==============================================
%============================================================================
\subsection{Perturbation}
Consider only an arbitrary scalar perturbation to the background FLRW metric, which is expressed by
\begin{eqnarray}\label{IV.18}
ds^2 & = & -(1+2A)dt^2 - 2a^2(t)\nabla_i B dx^i dt \nonumber \\
 & & + a^2(t)\Big[(1-2\psi)\delta_{ij} + 2 \nabla_i\nabla_j E  \Big]dx^i dx^j .
\end{eqnarray}
$\delta_{ij}$ is background spatial metric, and $\nabla_i$ is the covariant derivative with respect to this metric. The intrinsic curvature of the spatial hypersurface is expressed by the perturbation parameter $\psi$ as
\begin{equation}\label{IV.19}
^{3}R = {4 \over a^2} \; \nabla^2 \psi ,
\end{equation}
where $\psi$ is usually named the curvature perturbation. The curvature perturbation on the fixed-$t$ hypersurface is gauge dependent and under arbitrary coordinate transformation $t \rightarrow t + \delta t$; it transforms as $\psi \rightarrow \psi + H \delta t$. Any other scalar quantity such as energy density transforms as $\delta\rho \rightarrow \delta\rho - \dot\rho \delta t$ \cite{24,25,26}. The curvature perturbation for the uniform-density hypersurface ($\delta\rho=0$) is given by \cite{25}
\begin{equation}\label{IV.20}
\zeta = H \chi, \qquad \chi = {\psi \over H} + {\delta\rho \over \dot\rho} ,
\end{equation}
where $\chi$ is displacement between the uniform-density hypersurface and uniform-curvature hypersurface ($\psi=0$) \cite{24}. In contrast with $\psi$, $\zeta$ is a gauge-invariant perturbation parameter \cite{24,25,26}.\\
In addition to energy density, another component of the energy-momentum tensor is pressure. Perturbation of pressure splits into two parts as adiabatic and nonadiabatic, which is expressed as \cite{24}
\begin{equation}\nonumber
\delta p = c_s^2 \delta\rho + \dot{p}\Gamma ,
\end{equation}
where $c_s^2 = \dot{p}/\dot\rho$. The second term on the right-hand side is the nonadiabatic part, and
$\Gamma={\delta p \over \dot{p}} - {\delta\rho \over \dot\rho}$ is entropy perturbation, which is a
gauge-invariant parameter \cite{24}. For the energy conservation equation $n^\nu \nabla_\mu T^{\mu\nu}=0$, up to
first order of the energy perturbation, we have
\begin{equation}\label{IV.21}
\dot{\delta\rho} - 3H(\delta\rho + \delta p) - (\rho + p)\Big[ 3\dot\psi - \nabla^2(\sigma+\upsilon+B) \Big]=0 ,
\end{equation}
where $\sigma \equiv \dot{E} -B$ and $\nabla^i\upsilon$ is the perturbed 3-velocity of the fluid \cite{24,25}. In the uniform-density gauge (where $\delta\rho=0$ and $\zeta=\psi$) from Eq.\;(\ref{IV.21}) we have
\begin{equation}\label{IV.22}
\dot\zeta = {H \over (\rho+p)}\delta p_{nad} - {1 \over 3}\nabla^2(\sigma+\upsilon+B) .
\end{equation}
It should be noted that this result is independent of gravitational field equations, which is the main feature of this approach \cite{24}. From Eq.\;(\ref{IV.22}), it could be realized that $\zeta$ is not conserved generally. However, it is conserved if the pressure perturbation does not include a nonadiabatic part and the second part could be negligible. For a sufficiently large scale, the second part could be ignored, so \cite{27, 28}
\begin{equation}
\dot\zeta = {H \over (\rho+p)}\delta p_{nad} . \nonumber
\end{equation}
For the single scalar field case, the perturbation is adiabatic \cite{17}, so it could result that the curvature perturbation $\zeta$ is conserved on all superhorizon scales. The gauge-invariant perturbation $\zeta$ on a spatially flat hypersurface is given by
\begin{equation}\nonumber
\zeta \equiv \psi - {H \over \dot\rho}\delta\rho ,
\end{equation}
where for $ \psi=0 $ we have
\begin{equation}\label{IV.23}
\zeta = - {H \over \dot\rho}\delta\rho .
\end{equation}
According to the conservation equation $\dot\rho = -3H(\rho+p)$, energy density perturbation $\delta\rho = V' \delta\phi$ \cite{28a}, and Eq.\;(\ref{IV.23}), we have
\begin{equation}
\zeta =  {H \delta\phi \over \dot\phi}. \nonumber
\end{equation}
$\delta\phi$ on the large scale is given by $\delta\phi = H / 2\pi$. It is clearly seen that the form of perturbation parameter $\zeta$ is same as the standard one.
Using a power-law kinetic term and substituting Eqs.\;(\ref{III.11}) and (\ref{III.14}) into perturbation parameter $\zeta$, we then have
\begin{equation}\label{IV.24}
\zeta = \left({M_5^3 \over 4\pi n F_0\sqrt{2}}\right)^{-{2n \over 2n-1}} {-V'(\phi)V(\phi) \over 9nF_0 M_5^3}
\left( -V'(\phi) \over V(\phi) \right)^{-{2n \over 2n-1}}. \qquad
\end{equation}
In the following two subsections, we consider the results in more detail for two specific cases as $n=3/2$ and $n=2$.
Scalar fluctuations become seeds for cosmic microwave background (CMB) anisotropies or for large scale structure (LSS) formation. Therefore, by measuring the spectra of the CMB anisotropies and density distribution, the corresponding primordial perturbation could be determined. Besides scalar fluctuation, the inflationary scenario predicts tensor fluctuation, which is known as a gravitational wave, too. The produced tensor fluctuations induce a curved polarization in the CMB radiation and increase the overall amplitude of their anisotropies at a large scale. The physics of the early Universe could be specified by fitting the analytical results the of CMB and density spectra to corresponding observational data. At first, it was thought that the possible effects of primordial gravitational waves are not important and might be ignored. However, a few years ago, it was found out that the tensor fluctuations have an important role, and they should be more attended for determining best-fit values of the cosmological parameters from the CMB and LSS spectra \cite{1-2,2-1,2-2}. The imprint of tensor fluctuation on the CMB bring this idea to indirectly determine its contribution to power spectra by measuring CMB polarization \cite{2-1}. Such a contribution could be expressed by the $r$ quantity, which is known as tensor-to-scalar ratio and represents the relative amplitude of tensor-to-scalar fluctuation. Therefore, constraining $r$ is one of the main goals of the modern CMB survey. According to the current accuracy of observations, it is only possible to place a constant upper bound on the allowed range of $r$ \cite{2b-1}. Recent data from nine years of results of WMAP9 and South Pole Telescope (SPT) give the latest constraints of $r<0.13$, and $r<0.11$ at $95\%$ confidence level (C.L.) \cite{3-1,3-2,3-3,3-4}. Combining Planck's temperature anisotropy measurements with the WMAP large-angle polarization to constrain inflation, gives an upper limit $r<0.11$ in $95\%$ C.L. \cite{3-4,4-1}. Moreover, the BICEP2 collaboration completed three years of data, and their results expressed a constraint on the tensor-to-scalar ratio as $r=0.20^{+0.07}_{-0.05}$. \\
%============================================================================
%================= Subsection IV.B ==============================================
%============================================================================
\subsection{First case}
For the first case, we take $n=3/2$, which describes a noncanonical kinetic energy term as $F(X)=F_0X^{3/2}$,
where $F_0$ is a
constant with dimension $\rm{[M^{-2}]}$. Therefore, Eq.\;(\ref{III.14}) could be rewritten as
\begin{equation}\label{IV.25}
\dot\phi^2 = - {2\sqrt{2} \over 9F_0} \ {V' \over H}.
\end{equation}

\begin{figure}[ht]
\centering
\includegraphics[width=5cm]{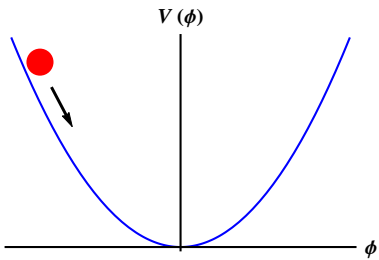}
\caption{Schematic of potential $V(\phi)=m^2\phi^2 / 2$.}\label{F01}
\end{figure}

\begin{figure}[ht] \centering
\includegraphics[width=5cm]{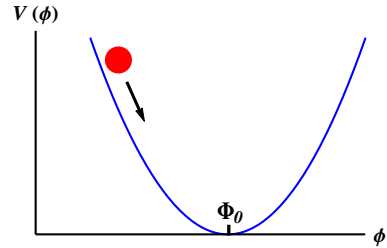}
\caption{Schematic of potential $V(\phi)=m^2(\phi-\Phi_0)^2 / 2$.}\label{F02}
\end{figure}

Since the right-hand side of equation is positive, to have a physical situation, we assume that $V'$ is negative.
This case could be acquired by taking the scalar field potential as $V(\phi)=m^2\phi^2 / 2$ so that during inflation the
scalar field approaches the minimum of its potential from left-hand side, Fig.\;\ref{F01}. Therefore, the value of scalar field is negative, and its time derivative $\dot\phi$ is positive.
%%%%%%%%%%%%%%%%%%%%%%%%%%%%%%%%%%%%%%%%%%%%%%%%%%%%%%
%%%%%%%%%%%%%%%%%%%%%%%%%%%%%%%%%%%%%%%%%%%%%%%%%%%%%%%
However, it is more common to take the scalar field as a positive value. Therefore, one can displace the minimum of the potential by assuming
 $V(\phi)= m^2(\Phi_0 - \phi)^2 / 2$, in which $\Phi_0$ is a positive constant, and during inflation, $\phi$ is always smaller than $\Phi_0$, Fig\;\ref{F02}. In this case, the scalar field is positive and growing by passing time, which might be more desirable. Also, its time derivative is positive as well. So, in this section, we take the potential as
\begin{equation}\label{IV.26}
V(\phi)= {1 \over 2}m^2(\Phi_0 - \phi)^2 .
\end{equation}
%%%%%%%%%%%%%%%%%%%%%%%%%%%%%%%%%%%%%%%%%%%%%%%%%%%%%%%%%
%%%%%%%%%%%%%%%%%%%%%%%%%%%%%%%%%%%%%%%%%%%%%%%%%%%%%%%%%
Substituting the above potential in the dynamical equations (\ref{III.11}) and (\ref{III.12}) leads to
\begin{eqnarray}
H & = & {2\pi \over 3M_5^3}\ m^2(\Phi_0 - \phi)^2 , \label{IV.27} \\
\dot{H} & = & - {4\pi^2 F_0  \over \sqrt{2}M5^6  } m^2 (\Phi_0 - \phi)^2 \dot\phi^3 . \label{IV.28}
\end{eqnarray}
Plugging Eqs.(\ref{IV.26}) and (\ref{IV.27}) into Eq. (\ref{IV.25}), the time derivative of scalar field is expressed by
\begin{equation}\label{IV.29}
\dot\phi^2 = {\sqrt{2}M_5^{3}  \over 3\pi F_0}{1\over (\Phi_0 - \phi)} .
\end{equation}
Based on Eq.\;(\ref{IV.16}), and the potential (\ref{IV.26}), the potential slow-roll parameter $\epsilon_V$ is obtained as
\begin{equation}\label{IV.30}
\epsilon_V \cong 0.37 \sqrt{M_5^9 \over F_0 m^4} \; {1 \over  (\Phi_0 - \phi)^{7/2} } .
\end{equation}
To have inflation, $\epsilon_V$ should be much smaller than unity, so that scalar field should satisfy condition $(\Phi_0 - \phi)^{7/2} \gg 0.37 m^{2}\sqrt{M_5^{9}/ F_0}$.
Inflation ends when $\ddot{a}=0$, which is equivalent to $\epsilon_V=1$. Therefore, the scalar field at the end of inflation is read as
\begin{equation}\label{IV.31}
(\Phi_0 - \phi_e)^{7/2} = 0.37 \sqrt{M_5^9 \over F_0 m^4} ,
\end{equation}
where $\phi_e$ denotes the value of the scalar field at the end of inflation. The other important parameter in studying inflation is the number of \textit{e}-folds. To get a successful inflation and overcome standard cosmology problems, we should have enough number of \textit{e}-folds. This parameter is defined by
\begin{eqnarray}
N & \equiv & \int^{t_e}_{t_i} H dt = \int^{\phi_e}_{\phi_i} {H \over \dot\phi} d\phi , \nonumber \\
 & = & \left( {9F_0 \over 2\sqrt{2}} \right)^{1/2} \ {1 \over m} \ \int_{\phi_i}^{\phi_e} {H^{3/2} \over (\Phi_0 - \phi)^{1/2}}d\phi . \label{IV.32}
\end{eqnarray}

By using Eq.\;(\ref{IV.27}), the number of \textit{e}-folds is given by
\begin{equation}\label{IV.33}
N = 1.54 \sqrt{F_0 m^4 \over M_5^9} \; \Big[ (\Phi_0-\phi_i)^{7/2} - (\Phi_0-\phi_e)^{7/2} \Big] .
\end{equation}
There should be about $55-65$ \textit{e}-folds to get rid of the standard cosmology problem. The initial scalar field is expressed by
\begin{equation}\label{IV.34}
(\Phi_0-\phi_i)^{7/2} = \sqrt{M_5^9 \over F_0 m^4} \left[ 0.37 + {N \over 1.54} \right] .
\end{equation}

Another potential slow-roll parameter is obtained by inserting the potential (\ref{IV.26}) into
Eq.(\ref{IV.17}) as
\begin{equation}\label{IV.35}
\eta_V(\phi) \cong 0.092 \sqrt{M_5^9 \over F_0 m^4} \ {1 \over (\Phi_0-\phi)^{7/2}} .
\end{equation}
It can be seen that the general forms of two potential slow-roll parameters are the same.
%============================================================================
%================= Subsubsection IV.B.1 ==========================================
%============================================================================
\subsubsection{Perturbation and observational constraints}
A brief review of perturbation was presented in the previous subsection. Then, by substituting Eq.\;(\ref{IV.34}) into  Eq.\;(\ref{IV.24}), the gauge-invariant perturbation $ \zeta $ is derived as
\begin{equation}\label{IV.36}
\zeta_i \cong {1.80 \; \alpha^{9/7} \over f^{1/7}} \; \left( {m \over  M_5} \right)^{10/7} ,
\end{equation}
where the dimensionless constant $f$ is a redefinition of $F_0$ and given by
\begin{equation}\label{IV.37}
f =F_0 M_5^{2} ;  \qquad \alpha = \left( 0.37 + {N \over 1.54} \right) .
\end{equation}
The amplitude of gauge-invariant perturbation is denoted by $A_s^2= 4 \zeta^2 /25$ \cite{17}, which is
obtained from observational data. Based on COBE and Planck data, one can derive a constraint on the free parameters
of the model. The results are expressed in Table.\ref{Tab01} for three different values of $N$.
%********************************************************************
%********************************************************************
\begin{table}[ht]
  \centering
  \begin{tabular}{lp{2cm}p{2cm}p{2cm}}
     \toprule[1.5pt] \\ [-2mm]
     % after \\: \hline or \cline{col1-col2} \cline{col3-col4} ...
         $N$     & $\qquad 55$ \qquad  & $\qquad 60$ \qquad  & $\qquad 65$ \\[0.5mm]
              \midrule[1pt] \\[-2mm]
     $\alpha$ \qquad\qquad & \ \ \ \ $36.08$  \qquad & \ \ \ \ $39.33$ \qquad & \ \ \ \ $42.58$  \\[1.3mm]
     $\tilde{m}_{\rm Pl}$ \qquad\qquad  & $4.66\times 10^{-5}$ \qquad  & $4.31\times 10^{-5}$ \qquad & $4.02\times 10^{-5}$  \\[1.3mm]
     $\tilde{m}_{\rm Co}$ \qquad\qquad  & $4.87\times 10^{-5}$ \qquad  & $4.50\times 10^{-5}$ \qquad & $4.19\times 10^{-5}$ \\[1.3mm]
     \bottomrule[1.5pt]
   \end{tabular}
  \caption{\footnotesize The constraint on the mass is obtained from Planck2013 data. The parameter $\alpha$ is defined from Eq.\;(\ref{IV.37}). $\tilde{m}$ is defined as $\tilde{m}=m/f^{0.1}M_5$, and the lower indices Pl and Co, respectively, indicate that the value of $\tilde{m}$ is obtained based on Planck-2013 and COBE data. The constraint on $\bar{m}$ is obtained for three different values of the number of \textit{e}-folds. The aAmplitude of scalar perturbation is estimated as $\ln\big(10^{10}A_s^2\big)=3.097$ from Planck data and $A_s=2\times 10^{-5}$ from COBE data.}\label{Tab01}
\end{table}
%********************************************************************
%********************************************************************
The other observational parameter is spectral index $n_s$, which is defined as
\begin{equation}\label{IV.38}
n_s-1 = {d\ln(A_s^2) \over d\ln(k)} = 2\eta_V(\phi_i) - 5 \epsilon_V(\phi_i) .
\end{equation}
The relations have a little difference in the coefficient with respect to ordinary models of inflation, where we have $n_s-1= 2\eta - 6\epsilon$. The initial potential slow-roll parameters are expressed as
\begin{equation}
\epsilon_V(\phi_i)= {0.37 \over \alpha},  \qquad \eta_V(\phi_i)={0.092 \over \alpha} . \nonumber
\end{equation}
Tensor perturbations could be used to find out the free parameters of our model. They are bounded to the brane at a long wavelength \cite{16,17} and  have no coupling to matter up to first order. Then, the amplitude of tensor perturbation is expressed by $A_T^2 = 4H^2 / 25\pi M_4^2$, at the Hubble crossing \cite{17,29}. Tensor index spectra are defined as
$n_T = {d\ln(A_T^2) \over d\ln(k)}=-2\epsilon_V(\phi_i)$. In Table {\ref{Tab02}}, the values of slow-roll parameters and spectral indices are specified. Also, the values of spectral indices are compared with recent observational data.\\
%********************************************************************
%********************************************************************
\begin{table}[ht]
  \centering
  \begin{tabular}{lp{1.5cm}p{1.5cm}p{1.5cm}p{1.5cm}}
     \toprule[1.5pt]\\[-2mm]
     % after \\: \hline or \cline{col1-col2} \cline{col3-col4} ...
     $\ \  N$     & $\quad 55$ \qquad  & $\quad 60$ \qquad   & $\quad 65$ \qquad & ${\rm observed}$ \\[0.5mm]
              \midrule[1pt]\\[-2mm]
     $\ \ \epsilon _V$ \qquad \qquad & $0.0102$ \qquad & $0.0094$ \qquad & $0.0086$ \qquad &  \\[1.3mm]
     $\ \ \eta _V$ \qquad \qquad & $0.0025$ \qquad & $0.0023$ \qquad & $0.0021$ \qquad & \\[1.3mm]
     $\ \ n_s$ \qquad \qquad & $0.9538$ \qquad  & $0.9576$ \qquad & $0.9608$ \qquad & $0.9675$ \qquad \\[1.3mm]
     $-n_T$ \qquad \qquad & $0.0205$ \qquad  & $0.0188$ \qquad & $0.0173$ \qquad & $<0.016$ \qquad \\[1.3mm]
     \bottomrule[1.5pt]
   \end{tabular}
  \caption{\footnotesize Scalar and tensor spectral indices are shown for three different values of the number of \textit{e}-folds $N$. The observed value of $n_s$ is related to Planck data, and $n_T$ is related to ${\rm WMAP9+eCMB+BAO+H_0}$ data. In a comparison, it could be realized that the theoretical values of these parameters are in good consistency with observational data.}\label{Tab02}
\end{table}
%********************************************************************
%********************************************************************
\noindent Two free parameters of the model are determined using observational data. Equation (\ref{IV.36}) applies the first constraint. The other constraint is necessary to specify the free parameter. The ratio of the tensor perturbation amplitude to the scalar perturbation amplitude is denoted by $r = A_T^2 / A_s^2$. This relation gives the second constraint. Using them, the free parameters of the model are derived. Table {\ref{Tab03}} summarizes information about our model for three different values of the number of \textit{e}-folds. It is seen that the scalar field mass
is obtained at about $10^{-6}M_5$, which is much smaller than its corresponding value in standard chaotic inflation.
Moreover, the value of the kinetic energy during inflation is smaller than the potential energy, which is consistence with our first assumption.
%************************************************************
%************************************************************
\begin{table}[ht]
  \centering
  \begin{tabular}{@{}p{1.2cm}p{2.2cm}p{2.2cm}p{2.2cm}@{}}
    \toprule[1.5pt]\\[-2mm]
    % after \\: \hline or \cline{col1-col2} \cline{col3-col4} ...
      $N$    & $\qquad 55$ & $\qquad 60$ & $\qquad 65$ \\[0.5mm]
          \midrule[1pt]\\[-2mm]
    $\bar{m}$  \qquad\qquad    & $3.228 \times 10^{-6}$ \qquad & $2.961 \times 10^{-6}$ \qquad & $2.735 \times 10^{-6}$ \\[1.53mm]
    $\bar{\phi}_i$ \qquad\qquad & $1.738 \times 10^{5}$ \qquad  & $1.894 \times 10^{5}$ \qquad  & $2.051 \times 10^{5}$ \\[1.53mm]
    $\bar{\phi}_e$ \qquad\qquad & $6.239 \times 10^{5}$ \qquad  & $6.635 \times 10^{5}$ \qquad  & $7.022 \times 10^{5}$ \\[1.53mm]
    $\bar{V}_i$ \qquad\qquad   & $1.574 \times 10^{-1}$ \qquad & $1.574 \times 10^{-1}$ \qquad & $1.574 \times 10^{-1}$ \\[1.53mm]
    $\bar{V}_e$ \qquad\qquad   & $2.028 \times 10^{-2}$ \qquad & $1.930 \times 10^{-2}$ \qquad & $1.845 \times 10^{-2}$ \\[1.53mm]
    $\bar{F}_i$ \qquad\qquad  & $1.792 \times 10^{-4}$ \qquad & $1.644 \times 10^{-4}$ \qquad & $1.519 \times 10^{-4}$ \\[1.53mm]
    $\bar{F}_e$ \qquad\qquad   & $8.33 \times 10^{-4}$ \qquad & $7.935 \times 10^{-4}$ \qquad & $7.583 \times 10^{-4}$ \\[1.53mm]
    \bottomrule[1.5pt]
  \end{tabular}
  \caption{\footnotesize Summary information about the parameter of the model. $\bar{x}$ variables denote: $\bar{m}=m/M_5$, $\bar{\phi}_i=(\Phi_0-\phi_i)/M_5$, $\bar{\phi}_e=(\Phi_0-\phi_e)/M_5$, $\bar{V}_i=V_i/M_5^4$, $\bar{V}_e=V_e/M_5^4$, $\bar{F}_e=F_e/M_5^4$, and $\bar{F}_e=F_e/M_5^4$ (where $F$ stands for the kinetic energy density of the scalar field).}\label{Tab03}
\end{table}
%*****************************************************************
%*****************************************************************
\begin{table}[ht]
  \centering
  \begin{tabular}{p{1.5cm}||p{1cm}p{1.5cm}p{1.5cm}p{1.5cm}}
    \toprule[1.5pt] \\[-2mm]
    % after \\: \hline or \cline{col1-col2} \cline{col3-col4} ...
    $N$  &  & $\quad 55$ & $\quad 60$  & $\quad 65$ \\[0.5mm]
      \midrule[1.5pt] \\[-2mm]
    $\tau = 10^{4}$ \qquad   & \ $\bar{t} _i$ \qquad\quad & $0.02720$ & $0.02605$ & $0.02504$ \\[1.5mm]
                             & \ $\bar{t} _e$ \qquad\quad & $154.785$ & $170.384$ & $186.057$ \\[1.3mm]
    \hline \\[-2mm]
    $\tau = 10^{5}$ \qquad   & \ $\bar{t} _i$ \qquad\quad & $0.27205$ & $0.26058$ & $0.25045$ \\[1.5mm]
                             & \ $\bar{t} _e$ \qquad\quad & $155.029$ & $170.618$ & $186.282$ \\[1.3mm]
    \hline \\[-2mm]
    $\tau = 10^{6}$ \qquad   & \ $\bar{t} _i$ \qquad\quad & $2.72054$ & $2.60586$ & $2.50454$ \\[1.5mm]
                             & \ $\bar{t} _e$ \qquad\quad & $157.478$ & $172.963$ & $188.537$ \\[1.3mm]
    \hline \\[-2mm]
    $\tau = 10^{7}$ \qquad   & \ $\bar{t} _i$ \qquad\quad & $27.2054$ & $26.0586$ & $25.0454$ \\[1.5mm]
                             & \ $\bar{t} _e$ \qquad\quad & $181.963$ & $196.416$ & $211.071$ \\[1.3mm]
    \bottomrule[1.5pt]
  \end{tabular}
  \caption{The times of the beginning and end of inflation have been determined for three different values of constant $c_1$ and the number of \textit{e}-folds $N$. Also, the dimensionless parameter $\bar{t}$ is defined as $\bar{t}=t/t_5$. Numbers of this table are obtained using the results of Table \ref{Tab03}.}\label{Tab04}
\end{table}
%*****************************************************************
%*****************************************************************
%============================================================================
%================= Subsubsection IV.B.2 ==========================================
%============================================================================
\subsubsection{Inflation time period}
Inflation is an early time Universe evolution which happens in very short time period, and the Universe undergoes an extreme expansion. Taking the integrate of Eq.\;(\ref{IV.29}) leads to the following equation for the scalar field
\begin{equation}\label{IV.39}
\Phi_0 - \phi(t) = \left( C_1 - 0.58 \; {M_5^{5/2} \over f^{1/2}} \; t \; \right)^{2/3}.
\end{equation}
$C_1$ is the constant of integration. One can rearrange the above equation to get the expression for time as
\begin{equation}\label{IV.40}
t = 1.72 \; f^{1/2} \; \Big( c_1 - \bar{\phi}^{3/2} \Big) \; t_5,
\end{equation}
in which $t_5$ is the five-dimensional planck time, $t_5=M_5^{-1}$, and the constant $c_1$ is redefined as
$C_1 = c_1 M_5^{3/2}$. Note that the time $t$ should be positive to have a physical situation; therefore, the constant
parameter $c_1$ should be larger than $\phi_i^{3/2}$. Without loss of generality, one can define this constant parameter as $c_1=\phi_i^{3/2} + \tau$.
Substituting the initial and final values of the scalar field from Table. \ref{Tab03} into Eq.(\ref{IV.40}),
the beginning and end inflation times are acquired for three different values of $\tau_1$,
which are displayed in Table {\ref{Tab04}}.
It is realized that, for any value of $\tau$, inflation begins earlier and ends latter for a bigger value of $N$.
The time period for $N=55, 60$, and $65$ are, respectively, obtained as $\triangle \bar{t}=154.757, 170.357$, and $186.032$.

%============================================================================
%================= Subsection IV.C ==============================================
%============================================================================
\subsection{Second case}
In this case, we take $n=2$ describing a noncanonical kinetic energy as $F=F_0X^2$. A chaotic potential $V(\phi)=m^2\phi^2 / 2$ is assumed for the scalar field. Therefore, Eq.\;(\ref{III.14}) is given by
\begin{equation}\label{IV.41}
\dot\phi^3 = -{V' \over 3F_0H} = -{m^2\phi \over 3F_0H} .
\end{equation}

\begin{figure}[ht]
\centering
\includegraphics[width=5cm]{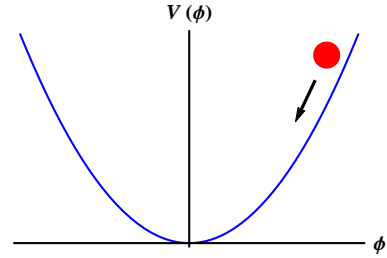}
\caption{Schematic of potential $V(\phi)= m^2\phi^2 / 2$ with negative $\dot\phi$.}\label{F03}
\end{figure}

In contrast with the previous case, the equation could be well justified by taking negative $\dot\phi$ and the positive scalar field value; refer to Fig. \ref{F03} (this situation occurs in the usual chaotic inflation).
%%%%%%%%%%%%%%%%%%%%%%%%%%%%%%%%%%%%%%%%%%%%%%%%%%%%%%%%%%%%%%%%%%%
%%%%%%%%%%%%%%%%%%%%%%%%%%%%%%%%%%%%%%%%%%%%%%%%%%%%%%%%%%%%%%%%%%%
\noindent So, we go further with this potential. Substituting this potential in the dynamical equations (\ref{III.11}) and (\ref{III.12}), one arrives at
\begin{eqnarray}
H & = & {2\pi \over 3M_5^3} m^2 \phi^2 , \label{IV.42}\\
\dot{H} & = & -{8\pi^2 \over 3M_5^6} F_0 m^2 \phi^2 \dot\phi^4 , \label{IV.43}
\end{eqnarray}
and the slow-roll parameters (\ref{IV.16}) and (\ref{IV.17}) could be reobtained as
\begin{eqnarray}
\epsilon_V &\cong& 0.52 \; {M_5^4 \over F_0^{1/3} m^2} \; {1 \over \phi^{10/3}} , \label{IV.44}\\
\eta_V &=& {V'' \over 9F_0H^2\dot\phi^2} \cong 0.086 \; {M_5^4 \over F_0^{1/3}m^2} \; {1 \over \phi^{10/3}} . \label{IV.45}
\end{eqnarray}
Inflation occurs for $\epsilon_V \ll 1$, which corresponds to scalar field values in the range $\phi^{10/3} \gg 0.52 \; M_5^4 / F_0^{1/3} m^2$. And, inflation ends when $\epsilon_V=1$, describing $\ddot{a}=0$ for the Universe, which corresponds to the following scalar field:
\begin{equation}\label{IV.46}
\phi_e^{10/3} = 0.52 \; {M_5^4 \over F_0^{1/3} m^2} .
\end{equation}
To overcome the standard big bang cosmology problem, we should have enough of a mount of inflation in early times of the Universe evolution. The number of \textit{e}-folds is obtained as
\begin{eqnarray}
N & \equiv & \int^{t_e}_{t_i} H dt = \int^{\phi_e}_{\phi_i} {H \over \dot\phi} d\phi , \nonumber \\
 & = & 1.16 \; {F_0^{1/3} m^2 \over M_5^4} \ \Big( \phi_i^{10/3} - \phi_e^{10/3} \Big) . \label{IV.47}
\end{eqnarray}

To get the desirable situation, we should have $N=55-65$. From Eqs.\;(\ref{IV.46}) and (\ref{IV.47}), the initial scalar field is obtained as
\begin{equation}\label{IV.48}
\phi_i^{10/3} = \Big( 0.52 + {N \over 1.16} \Big) \; {M_5^4 \over F_0^{1/3}m^2} .
\end{equation}
%============================================================================
%================= Subsubsection IV.C.1 ==========================================
%============================================================================
\subsubsection{Perturbation and observational constraint}
A brief review of perturbation parameters was presented in the previous section. Then, the gauge-invariant curvature perturbation $\zeta$ for this case is derived as
\begin{equation}\label{IV.49}
\zeta = {H\delta\phi \over \dot\phi} ;
\end{equation}
the general form of $\zeta$ is similar to the standard inflationary models. Using the Friedmann equation (\ref{IV.42}) and initial value of the scalar field (\ref{IV.48}), $\zeta$ is estimated as
\begin{equation}\label{IV.50}
\zeta_i \cong 1.29 \; {\beta^{13/10} \over f^{1/10}} \; \left( {m \over M_5} \right)^{7/5} ,
\end{equation}
where, without loss of generality, we defined
\begin{equation}\label{IV.51}
f=F_0 M_5^4; \qquad \beta = 0.52 + {N \over 1.16} .
\end{equation}
Note that $f$ is a dimensionless constant. To justify the dimension of the kinetic term, the constant $F_0$ should have dimension ${\rm [M^{-4}]}$, which is expressed in Table \ref{Tab05}.
%********************************************************************
%********************************************************************
\begin{table}[ht]
  \centering
  \begin{tabular}{lp{2cm}p{2cm}p{2cm}}
     \toprule[1.5pt]\\[-2mm]
     % after \\: \hline or \cline{col1-col2} \cline{col3-col4} ...
          $N$    & $\qquad 55$ \qquad  & $\qquad 60$ \qquad  & $\qquad 65$ \\[0.5mm]
              \midrule[1pt]\\[-2mm]
     $\beta$ \qquad\qquad & \ \ \ \ $47.93$ \qquad & \ \ \ \ $52.24$ \qquad & \ \ \ \ $56.55$  \\[1.3mm]
     $\tilde{m}_{\rm Pl}$ \qquad\qquad  & $3.58\times 10^{-5}$ \qquad  & $3.30\times 10^{-5}$ \qquad & $3.07\times 10^{-5}$  \\[1.3mm]
     $\tilde{m}_{\rm Co}$ \qquad\qquad  & $3.73\times 10^{-5}$ \qquad  & $3.44\times 10^{-5}$ \qquad & $3.20\times 10^{-5}$ \\[1.3mm]
     \bottomrule[1.5pt]
   \end{tabular}
  \caption{\footnotesize The constraint on the mass is obtained from Planck2013 data. The parameter $\beta$ is defined from Eq.\;(\ref{IV.51}). $\tilde{m}$ is defined as $\tilde{m}=m/f^{1/14}M_5$, and the lower indices Pl and Co, respectively, indicate that the value of $\tilde{m}$ is obtained based on Planck-2013 and COBE data. The constraint on $\bar{m}$ is obtained for three different values of the number of \textit{e}-folds. The amplitude of scalar perturbation is estimated as $\ln\big(10^{10}A_s^2\big)=3.097$ from Planck data and $A_s=2\times 10^{-5}$ from COBE data.}\label{Tab05}
\end{table}
%********************************************************************
%********************************************************************
\noindent  As in the previous case, the constraint of the scalar field mass is obtained for three different values of $N$. The results show that $m$ is about $10^{-5}\; f^{0.1}M_5$. \\
The spectral index of the scalar perturbation that presents the scale dependence of the scalar perturbation amplitude is given by
\begin{equation}\label{IV.57}
n_s-1 = {d\ln(A_s^2) \over d\ln(k)} = 2\eta_V(\phi_i) - 13\epsilon_V(\phi_i)/3 ,
\end{equation}
for which $\eta_V(\phi_i)$ and $\epsilon_V(\phi_i)$ indicate the potential slow-roll parameters for the initial value of the scalar field and are given by
\begin{equation}\label{IV.58}
\epsilon_V(\phi_i) = {0.52 \over \beta}, \qquad \eta_V(\phi_i) = {0.086 \over \beta} .
\end{equation}
The amplitude of the tensor perturbation is given by $A_T^2 = 4H^2 / 25\pi M_4^2$, and the tensor spectral index is defined as $n_T=-2\epsilon$. The slow-rolling parameters and spectral indices quantities are expressed in Table {\ref{Tab06}} and compared with observational data.
%********************************************************************
%********************************************************************
\begin{table}[ht]
  \centering
  \begin{tabular}{lp{1.5cm}p{1.5cm}p{1.5cm}p{1.5cm}}
     \toprule[1.5pt]\\[-2mm]
     % after \\: \hline or \cline{col1-col2} \cline{col3-col4} ...
       $\ \ N$       & $\quad 55$ \qquad  & $\quad 60$ \qquad   & $\quad 65$ \qquad & ${\rm Observed}$ \\[0.5mm]
              \midrule[1pt]\\[-2mm]
     $\ \ \epsilon _V$ \qquad \qquad & $0.0108$ \qquad & $0.0099$ \qquad & $0.0091$ \qquad &  \\[1.3mm]
     $\ \ \eta _V$ \qquad \qquad & $0.0018$ \qquad & $0.0016$ \qquad & $0.0015$ \qquad & \\[1.3mm]
     $\ \ n_s$ \qquad \qquad & $0.9565$ \qquad  & $0.9601$ \qquad & $0.9631$ \qquad & $0.9675$ \qquad \\[1.3mm]
     $-n_T$ \qquad \qquad & $0.0216$ \qquad  & $0.0199$ \qquad & $0.0183$ \qquad & $<0.016$ \qquad \\[1.3mm]
     \bottomrule[1.5pt]
   \end{tabular}
  \caption{\footnotesize Scalar and tensor spectral indices are shown for three different values of the number of \textit{e}-folds $N$. The observed value of $n_s$ is related to Planck data, and $n_T$ is related to ${\rm WMAP9+eCMB+BAO+H_0}$ data. In a comparison, it could be realized that the theoretical values of these parameters are in good consistency with observational data.}\label{Tab06}
\end{table}
%********************************************************************
%********************************************************************
\begin{table}[ht]
  \centering
  \begin{tabular}{@{}p{1.2cm}p{2.2cm}p{2.2cm}p{2.2cm}@{}}
    \toprule[1.5pt]\\[-2mm]
    % after \\: \hline or \cline{col1-col2} \cline{col3-col4} ...
      $N$    & $\qquad 55$ & $\qquad 60$ & $\qquad 65$ \\[0.5mm]
          \midrule[1pt]\\[-2mm]
    $\bar{m}$  \qquad\qquad   & $3.865 \times 10^{-6}$ \qquad & $3.546 \times 10^{-6}$ \qquad & $3.276 \times 10^{-6}$ \\[1.53mm]
    $\bar{\phi}_i$ \qquad\qquad & $1.273 \times 10^{5}$ \qquad & $1.387 \times 10^{5}$ \qquad  & $1.501 \times 10^{5}$ \\[1.53mm]
    $\bar{\phi}_e$ \qquad\qquad & $3.986 \times 10^{4}$ \qquad & $4.234 \times 10^{4}$ \qquad  & $4.476 \times 10^{4}$ \\[1.53mm]
    $\bar{V}_i$ \qquad\qquad  & $1.210 \times 10^{-1}$ \qquad & $1.210 \times 10^{-1}$ \qquad & $1.210 \times 10^{-1}$ \\[1.53mm]
    $\bar{V}_e$ \qquad\qquad   & $1.187 \times 10^{-2}$ \qquad & $1.127 \times 10^{-2}$ \qquad & $1.075 \times 10^{-2}$ \\[1.53mm]
    $\bar{F}_i$ \qquad\qquad  & $1.089 \times 10^{-4}$ \qquad & $9.994 \times 10^{-5}$ \qquad & $9.232 \times 10^{-5}$ \\[1.53mm]
    $\bar{F}_e$ \qquad\qquad   & $9.755 \times 10^{-3}$ \qquad & $9.304 \times 10^{-3}$ \qquad & $8.907 \times 10^{-3}$ \\[1.53mm]
    \bottomrule[1.5pt]
  \end{tabular}
  \caption{\footnotesize Summary of information about the parameter of the model. $\bar{x}$ variables denote: $\bar{m}=m/M_5$, $\bar{\phi}_i=(\Phi_0-\phi_i)/M_5$, $\bar{\phi}_e=(\Phi_0-\phi_e)/M_5$, $\bar{V}_i=V_i/M_5^4$, $\bar{V}_e=V_e/M_5^4$, $\bar{F}_e=F_e/M_5^4$, and $\bar{F}_e=F_e/M_5^4$ (where $F$ stands for the kinetic energy density of the scalar field).}\label{Tab07}
\end{table}
%************************************************************
%************************************************************
Scalar perturbation with the tensor perturbation is used to determined the free parameters of the model, namely, $m$ and $f$. Then, the value of scalar field and potential energy are specified at the beginning and end of inflation. Table {\ref{Tab07}} summarizes this information.
%*****************************************************************
%*****************************************************************
\begin{table}[ht]
  \centering
  \begin{tabular}{p{1.5cm}||p{1cm}p{1.5cm}p{1.5cm}p{1.5cm}}
    \toprule[1.5pt] \\[-2mm]
    % after \\: \hline or \cline{col1-col2} \cline{col3-col4} ...
    $N$  &  & $\quad 55$ & $\quad 60$  & $\quad 65$ \\[0.5mm]
      \midrule[1.5pt] \\[-2mm]
    $\tau_2 = 10^{3}$ \qquad   & \ $\bar{t} _i$ \qquad\quad & $0.04018$ & $0.03905$ & $0.03803$ \\[1.5mm]
                               & \ $\bar{t} _e$ \qquad\quad & $202.645$ & $222.882$ & $243.214$ \\[1.3mm]
    \hline \\[-2mm]
    $\tau_2 = 10^{4}$ \qquad   & \ $\bar{t} _i$ \qquad\quad & $0.40188$ & $0.39050$ & $0.38032$ \\[1.5mm]
                               & \ $\bar{t} _e$ \qquad\quad & $203.006$ & $223.233$ & $243.556$ \\[1.3mm]
    \hline \\[-2mm]
    $\tau_2 = 10^{5}$ \qquad   & \ $\bar{t} _i$ \qquad\quad & $4.01880$ & $3.90509$ & $3.80325$ \\[1.5mm]
                               & \ $\bar{t} _e$ \qquad\quad & $206.623$ & $226.748$ & $246.979$ \\[1.3mm]
    \hline \\[-2mm]
    $\tau_2 = 10^{6}$ \qquad   & \ $\bar{t} _i$ \qquad\quad & $40.1880$ & $39.0509$ & $38.0325$ \\[1.5mm]
                               & \ $\bar{t} _e$ \qquad\quad & $242.792$ & $261.894$ & $281.208$ \\[1.3mm]
    \bottomrule[1.5pt]
  \end{tabular}
  \caption{The times of the beginning and end of inflation have been determined for three different values of constant $c_1$ and number of \textit{e}-folds $N$. Also, the dimensionless parameter $\bar{t}$ is defined as $\bar{t}=t/t_5$. Numbers of this table are obtained using the results of Table \ref{Tab07}.}\label{Tab08}
\end{table}
%*****************************************************************
%*****************************************************************
%============================================================================
%================= Subsubsection IV.C.2 ==========================================
%============================================================================
\subsubsection{Inflation time period}
Taking the integrate of Eq.\;(\ref{IV.41}) leads one to the expression for the scalar field as
\begin{equation}
\phi(t) = \left( C_2 - {0.72 \; M_5^{7/3} \over f^{1/3}} \; t \right)^{3/4},
\end{equation}
where $C_2$ is the constant of integration. By rearranging the above relation, one can get an expression for time that reads as
\begin{equation}
t = 1.39 \; f^{1/3} \; \Big( c_2 - \bar{\phi}^{4/3} \Big) \; t_5,
\end{equation}
where $t_5$ is the five-dimensional Planck time and $t_5=M_5^{-1}$. The constant $c_2$ is a redefinition of $C_2$ as
$C_2 = c_2 M_5^{4/3}$, which should be bigger than $\phi_i^{4/3}$ to give a positive value for time.
As in the previous case, the constant parameter $c_2$ could be expressed by a new parameter $\tau_2$.
The time period is displayed in Table {\ref{Tab08}} for three different values of $\tau_2$.
Almost the same results in the previous case are achieved. Inflation occurs earlier, and ends later by increasing the number of \textit{e}-folds. However, in comparison, the time period of inflation is larger than in the previous case in which for $N=55, 60$ and $65$, respectively, there is $\triangle \bar{t}= 202.604, 222.843$, and $243.176$. \\

%==========================================================================
%==========================================================================
%====================  Section V =============================================
%==========================================================================
%==========================================================================
\section{BICEP2 Result}
BICEP2 is the second generation of the background imaging cosmic extragalactic polarization (BICEP) instrument, which is placed at the South Pole. It is designed to measure the polarization of the CMB on an angular scale of $1$ to $5$, near the peak of the B-mode polarization signature of primordial gravitational waves from cosmic inflation. Measuring B-modes requires a dramatic improvement in sensitivity combined with exquisite control of systematics. The successful strategy of BICEP achieved the most sensitive limit on B-modes. Besides that, BICEP2 took advantage of its new detector design to get more detected than the first BICEP. The most simple and economical remaining interpretation of the B-modes signal is that it is due to tensor modes. Therefore, by detecting B modes, it is possible to set a constraint on the tensor-to-scalar ratio. BICEP2 has completed three years of observation (2010-2012). The report given in March 2014 stated that BICEP2 hase detected B-modes from gravitational waves of the early Universe. An announcement was made on March 17, 2014, that BICEP2 has detected B modes at the level of $r=0.20^{+0.07}_{-0.05}$ (for more details about the results and experiment, refer to Refs. \cite{36,37}). \\
According to BICEP2 results, the consequence of the presented model could be revisited. In the following table, the tensor-to-scalar ratio is taken as $r=0.20$, and some main parameters of the model are determined. From Table \ref{Tab09}, it is realized that the scalar field mass gets smaller values than two previous cases. The spectral indices remain as in the previouscases, which expressed that in this model the parameter $r$ has no effect on spectral indices. \\
%%%%%%%%%%%%%%%%%%%%%%%%%%%%%%%%%%%%%%%%%%%%%%%%%%%%%%%%
\begin{widetext}

\begin{table}[ht]
\centering
\begin{tabular}{@{}p{2cm}p{1cm}p{1cm}p{1cm}p{0.5cm}p{1cm}p{1cm}p{1cm}p{0.5cm}lll@{}}
  \toprule[1.5pt]\\[-2mm]
           & \multicolumn{3}{c}{$n=3/2$} & \qquad &  \multicolumn{3}{c}{$n=2$} & \qquad &  \multicolumn{3}{c}{observed} \\[0.5mm]
 \cline{2-4} \cline{6-8} \cline{10-12} \\[-2mm]
    $\quad N$ &  $\quad 55$ & $\quad 60$  & $\quad 65$  & \qquad & $\quad 55$ & $\quad 60$  & $\quad 65$ & \qquad &   & &                               \\[0.5mm]
     \midrule[1pt] \\[-2mm]

    $\bar{m}$ \tiny{($\times10^5$)} \quad    &  $0.2714$ \qquad & $0.2490$ \qquad & $0.2300$  & \qquad & $0.3250$  & $0.2982$  & $0.2755$  & \qquad & & &  \\[1.3mm]

     $n_s$ \qquad   &  $0.9538$  & $0.9576$ & $0.9608$  & \qquad & $0.9565$ & $0.9601$ & $0.9631$ & \qquad & & $0.9675$ &    \\[1.3mm]

     $-n_T$ \qquad  &  $0.0205$  & $0.0188$ & $0.0173$  & \qquad & $0.0216$   & $0.0199$  & $0.0183$ & \qquad & & $<0.016$ &    \\[1.3mm]

 \bottomrule[1.5pt]
\end{tabular}
\caption{Some parameters of the model based on BICEP2 results.}\label{Tab09}
\end{table}

\end{widetext}
%%%%%%%%%%%%%%%%%%%%%%%%%%%%%%%%%%%%%%%%%%%%%%%%%%%%%%%%%%

%==========================================================================
%==========================================================================
%====================  Section VI =============================================
%==========================================================================
%==========================================================================
\section{Conclusion}
In this paper, we have studied one of the most important evolutions of the Universe in the early times, named inflation, by using the RS-II braneworld gravity model. As in standard cosmology, it is assumed that in this era the Universe is dominated by a dynamical scalar field that imposes a quasi-de Sitter expansion. The usual conditions for such an expansion are that the rate of $H$ and $\dot\phi$ during a Hubble time should be much smaller than unity. These conditions are known as \textit{slow-roll conditions}, and this type of inflation is named slow-roll inflation. Up to now, different models of inflation have been proposed, and it could be said that chaotic inflation, with all advantages and defects, is one of the most successful inflationary scenarios that have received cosmologists' attention. The inflationary scenario in RS-II braneworld gravity comes to interesting consequence in the sense that in the same period of time we could have more expansion for the Universe. In this model of brane gravity, the presence of a quadratic of energy density in the Friedmann equation, which plays an important role in the high-energy regime, is the reason for such expansion. Therefore, the scenario of inflation in braneworld gravity is effective for a wider range of potentials.\\
During this work, the situation was somehow different from the usual case so that the scalar field possessed a noncanonical kinetic energy density. These types of scalar fields are known as noncanonical scalar fields with Lagrangian $\mathcal{L}=F(X)-V(\phi)$ and are a subclass of the general case with Lagrangian $\mathcal{L}=f(\phi)F(X)-V(\phi)$. $F(X)$ is an arbitrary function of $X=\dot\phi^2/2$. In this paper, $F(X)$ was picked out as a power-law function, $F(X)=F_0X^n$, and the potential was selected as a common form of the chaotic potential, $V(\phi)=m^2\phi^2/2$. Therefore, in brief, it could be said that the model concerned a chaotic inflation in RS-II braneworld gravity driven by a noncanonical scalar field. \\
The model was studied in more detail for two specific cases as $n=3/2, 2$. Then, the slow-roll conditions were applied, and some important parameters of inflationary scenario such as $m$, $n_s$, $n_T$ \ldots were derived and compared with observational data. A brief summary of these results is presented in Table \ref{TCon}.\\
%%%%%%%%%%%%%%%%%%%%%%%%%%%%%%%%%%%%%%%%%%%%%%%%%%%%%%%%
\begin{widetext}

\begin{table}[ht]
\centering
\begin{tabular}{@{}p{2cm}p{1cm}p{1cm}p{1cm}p{0.5cm}p{1cm}p{1cm}p{1cm}p{0.5cm}lll@{}}
  \toprule[1.5pt]\\[-2mm]
           & \multicolumn{3}{c}{$n=3/2$} & \qquad &  \multicolumn{3}{c}{$n=2$} & \qquad &  \multicolumn{3}{c}{observed} \\[0.5mm]
 \cline{2-4} \cline{6-8} \cline{10-12} \\[-2mm]
    $\quad N$ &  $\quad 55$ & $\quad 60$  & $\quad 65$  & \qquad & $\quad 55$ & $\quad 60$  & $\quad 65$ & \qquad &   & & \\[0.5mm]
     \midrule[1pt] \\[-2mm]

    $\bar{m}$ \tiny{($\times10^5$)} \quad    &  $0.3228$ \qquad & $0.2961$ \qquad & $0.2735$  & \qquad & $0.339$  & $0.311$  & $0.287$  & \qquad & & &  \\[1.3mm]

    $\bar{m}_{B}$ \tiny{($\times10^5$)} \quad    &  $0.2714$ \qquad & $0.2490$ \qquad & $0.2300$  & \qquad & $0.3250$  & $0.2982$  & $0.2755$  & \qquad & & &  \\[2mm]

     $n_s$ \qquad   &  $0.9538$  & $0.9576$ & $0.9608$  & \qquad & $0.9565$ & $0.9601$ & $0.9631$ & \qquad & & $0.9675$ &    \\[1.3mm]

     $-n_T$ \qquad  &  $0.0205$  & $0.0188$ & $0.0173$  & \qquad & $0.0216$   & $0.0199$  & $0.0183$ & \qquad & & $<0.016$ &    \\[1.3mm]

 \bottomrule[1.5pt]
\end{tabular}
\caption{A brief results of the model}\label{TCon}
\end{table}

\end{widetext}
%%%%%%%%%%%%%%%%%%%%%%%%%%%%%%%%%%%%%%%%%%%%%%%%%%%%%%%%%%
For both cases, the scalar field was obtained of order $10^{-6}$ of the five-dimensional Planck mass ($m \propto 10^{-6}M_5$); however, for $n=2$, it is slightly larger than the case $n=3/2$, Whereas the mass of scalar field in the standard inflationary scenario was derived as $10^{-6}M_4$, which is much larger than our predicted value. Also, in comparison to Ref \cite{17}, our results were about ten times smaller.\\
Spectral indices were obtained independently of the scalar field mass. For both cases, a larger value of the number of \textit{e}-folds resulted in more appropriate values $n_s$ and $n_T$, which are in acceptable consistence with observational data. The case $n=2$ had a better result for the scalar spectra. index. which is closer to observational data than the case $n=3/2$; however, the situation was the reverse for the tensor spectral index, in which the case $n=3/2$ had a better result. It seems that by increasing $n$, the theoretical value of $n_s$ come closer to the observational value, but the theoretical value of $n_T$ went away from its observational value.\\
Then, the estimated values of scalar field the energy densities were obtained for both cases. It was shown that the potential energy density in the beginning of inflation is much bigger than the kinetic energy and that the potential part of the energy density could easily dominate the kinetic part. Inflation began, and the scalar field slowly came down to the minimum of its potential (describing a slow-rolling inflation). By passing time, the potential energy density decreases and kinetic energy density increases, and inflation ends at the time $t_e$. \\
The beginning and end of inflation times were calculated in the last part of each of the cases, and the results were prepared in Tables \ref{Tab04} and \ref{Tab08}. It was seen that, depending on free parameter $\tau$, inflation could occur after the five-dimensional Planck time $t_5$. Thr time period of inflation was computed for three different values of $N$, for both cases. It was found out that inflation lasts longer for a larger number of \textit{e}-folds $N$. The results showed that there is bigger time period for the case $n=2$. \\
 The BICEP2 collboration has detected the B modes from gravitational waves of the early Universe and found a constraint on the tensor-to-scalar ratio as $r=0.20^{+0.07}_{-0.05}$. By taking this result, some parameters of the model were revisited. Based on the BICEP2 result, it was shown that the scalar field mass gets smaller values than the two previous cases; however, the spectral indices remain the same.  \\
%==========================================================================
%==========================================================================
%====================  Section V =============================================
%==========================================================================
%==========================================================================
\section{ACKNOWLEDGMENT}
The authors thank the referee and editor for giving this chance to add the new observational results of BICEP2 to manuscript.

%=====================================================================
%======================= Reference ======================================
%=====================================================================

\end{document}